\documentclass[twocolumn,preprintnumbers,amsmath,amssymb,floatfix,prl]{revtex4}
\usepackage{graphicx}% Include figure files
\usepackage{bm}% bold math

\usepackage{graphics}
\usepackage{color}
\usepackage{hyperref}
\newcommand{\LIA}{\lambda_{\rm I}^{\rm A}}
\newcommand{\LIB}{\lambda_{\rm I}^{\rm B}}
\newcommand{\LR}{\lambda_{\rm R}}
\newcommand{\LPIAA}{\lambda_{\rm PIA}^{\rm A}}
\newcommand{\LPIAB}{\lambda_{\rm PIA}^{\rm B}}
\newcommand{\LPIA}{\lambda_{\rm PIA}}
\begin{document}

%\preprint{APS/123-QED}

\title{Trivial and inverted Dirac bands, and emergence of quantum spin Hall states in graphene on transition-metal dichalcogenides}

\author{Martin Gmitra, Denis Kochan, Petra H\"{o}gl, and Jaroslav Fabian}

\affiliation{Institute for Theoretical Physics, University of Regensburg, 93040 Regensburg, Germany\\
}

\begin{abstract}
Proximity orbital and spin-orbital effects of graphene on monolayer transition-metal dichalcogenides
(TMDCs) are investigated from first-principles. The Dirac band structure
of graphene is found to lie within the semiconducting gap of TMDCs for sulfides and selenides, while
it merges with the valence band for tellurides. In the former case the proximity-induced staggered
potential gaps and spin-orbit couplings (all on the meV scale) of the Dirac electrons are established by fitting to a phenomenological
effective Hamiltonian. While graphene on MoS$_2$, MoSe$_2$, and WS$_2$ has a topologically trivial band structure,
graphene on WSe$_2$ exhibits inverted bands. Using a realistic tight-binding model we find topologically protected helical edge states for graphene zigzag nanoribbons on WSe$_2$, demonstrating the quantum spin
Hall effect. This model also features ``half-topological states'', which are protected against
time-reversal disorder on one edge only.
\end{abstract}

\maketitle

There has recently been a strong push to find ways to enhance spin-orbit coupling in graphene~\cite{Han2014:NatNano} to enable spintronics applications~\cite{Zutic2004:RMP, Fabian2007:APS}. Decorating graphene with adatoms~\cite{Neto2009:PRL, Gmitra2013:PRL} has proven particularly promising, as demonstrated experimentally by the giant spin Hall effect signals~\cite{balakrishnan_colossal_2013, Avsar2014:NatComm}. In parallel, there have been intensive efforts to predict realistic graphene structures that would exhibit
the quantum spin (and anomalous) Hall effect~\cite{Qiao2010:PRB, Weeks2011:PRX, Zhang2012:PRL, qiao_quantum_2014}, introduced by Kane and Mele~\cite{Kane2005:PRL} as a precursor of topological insulators~\cite{bernevig_quantum_2006, konig_quantum_2007, zhang_topological_2009}.

Ideal for inducing a large proximity spin-orbit coupling in graphene would be a matching 2d insulating or semiconducting material to preserve the Dirac band structure at the Fermi level.
Hexagonal BN is a nice substrate for graphene, but it has a weak spin-orbit
coupling itself~\cite{Han2014:NatNano}, so the proximity effect is negligible. The next best candidates are two-dimensional transition-metal dichalcogenides (TMDCs) which are direct band-gap semiconductors~\cite{Mak2010:PRL, Kormanyos2015:2DM}. Graphene on TMDCs has already been grown~\cite{Lin2014:ACS, Lin2014:APL, Azizi2015:ACS} and investigated
for transport~\cite{Lu2014:PRL, Larentis2014:NL}
as well as considered for technological applications~\cite{Kumar2015:MT, Bertolazzi2013:ACSNano, Zhang2014:SREP, Roy2013:NatNanotech}. It was recently predicted that monolayer MoS$_2$ will induce a giant
spin-orbit coupling in graphene, of about 1~meV (compared to 10~$\mu$eV in pristine
graphene~\cite{Gmitra2009:PRB}). A recent experiment~\cite{Avsar2014:NatComm} on the
room temperature spin Hall effect in graphene on few layers of WS$_2$ found a large spin-orbit coupling,
about 17~meV, attributing it to defects in the thin WS$_2$, rather than to the genuine
proximity effect.

As the proximity spin-orbit coupling in graphene on TMDCs is expected to grow with the increasing
atomic number of the transition-metals, we here explore the whole family of TMDCs as potential
substrates for graphene. In most cases we find trivial Dirac cones, affected by the proximity
effects. But for graphene on WSe$_2$ we see a robust band inversion and emergent spin Hall
effect in the corresponding zigzag nanoribbons.

\begin{figure}[h!]
 \includegraphics[width=0.99\columnwidth]{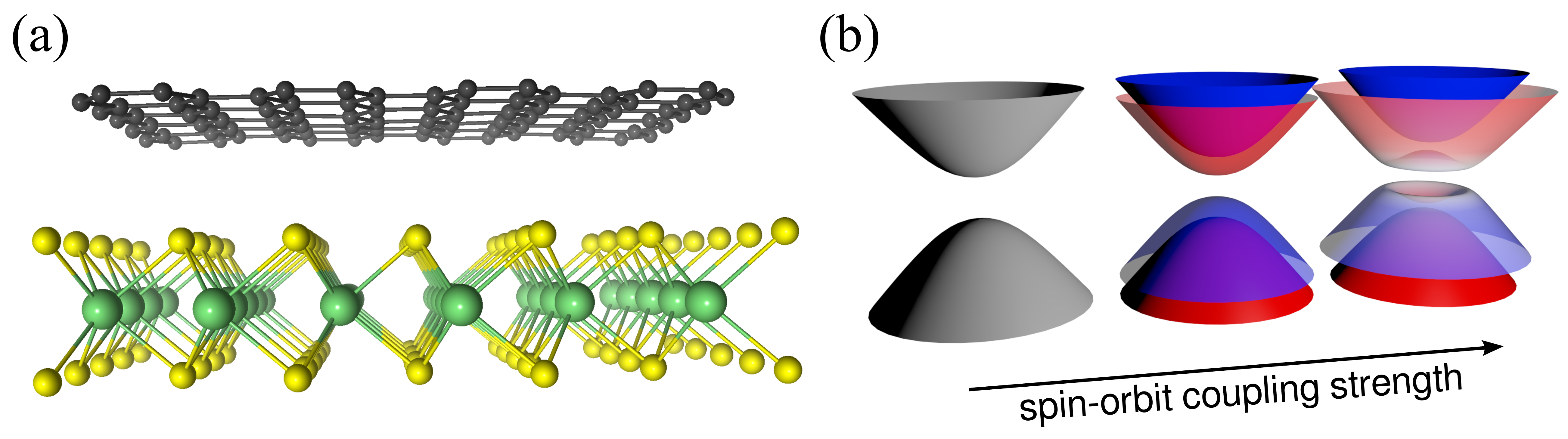}
 \caption{Sketch of
(a)~the atomic structure of graphene on a monolayer transition-metal dichalcogenide, and
(b)~evolution of spin-orbit coupling induced band structure topology
near the Dirac point of graphene in this hybrid structure. In the absence of
spin-orbit coupling the doubly spin degenerate bands are split by the proximity orbital gap.
As the proximity spin-orbit coupling is turned on, first the spin degeneracy is lifted, followed by the band inversion at large values of spin-orbit coupling. Red and blue colors indicate
the opposite spin projections along the transverse ($z$) axis.
 }\label{Fig:struct}
\end{figure}

More specifically, we report here on systematic first-principles calculations 
predicting that (i) graphene on MoS$_2$, WS$_2$, MoSe$_2$, WSe$_2$, MoTe$_2$, and WTe$_2$
monolayers [see Fig.~\ref{Fig:struct}(a) for the structure] preserves its linear-in-momentum band structure
%------------------------------------------------------------------------
%------------------------------------------------------------------------
within the TMDCs direct band gaps, shifting the Dirac point towards the valence bands
of TMDCs with increasing the atomic number of the chalcogen;
graphene on transition-metal tellurides
has the Dirac point merged with the TMDCs valence bands. (ii) The proximity spin-orbit
coupling increases with the atomic number of the transition metal. While the Dirac band structure in most
cases is conventional, (iii) graphene on WSe$_2$ exhibits a band inversion due to the
anticrossings of graphene's conduction and valence bands that are spin-polarized in the opposite directions. The evolution
of the graphene band structure from pristine, through trivial proximity, and to nontrivial band inversion,
as the proximity spin-orbit coupling increases, is sketched in Fig.~\ref{Fig:struct}(b).
Using realistic tight-binding modeling of the proximity-induced orbital and spin-orbital effects
in graphene on WSe$_2$ we further show that (iv) zigzag graphene nanoribbons
in this structure have helical edge states inside the bulk gap, demonstrating the quantum
spin Hall effect. We also find that (v) states outside the gap exhibit a pronounced edge asymmetry, with
an odd number of pairs at one edge and even number of pairs at the other edge. We call
such states half-topological, as they are protected against time-reversal impurity scattering at one edge only.

\begin{figure}[h!]
 \includegraphics[width=0.99\columnwidth]{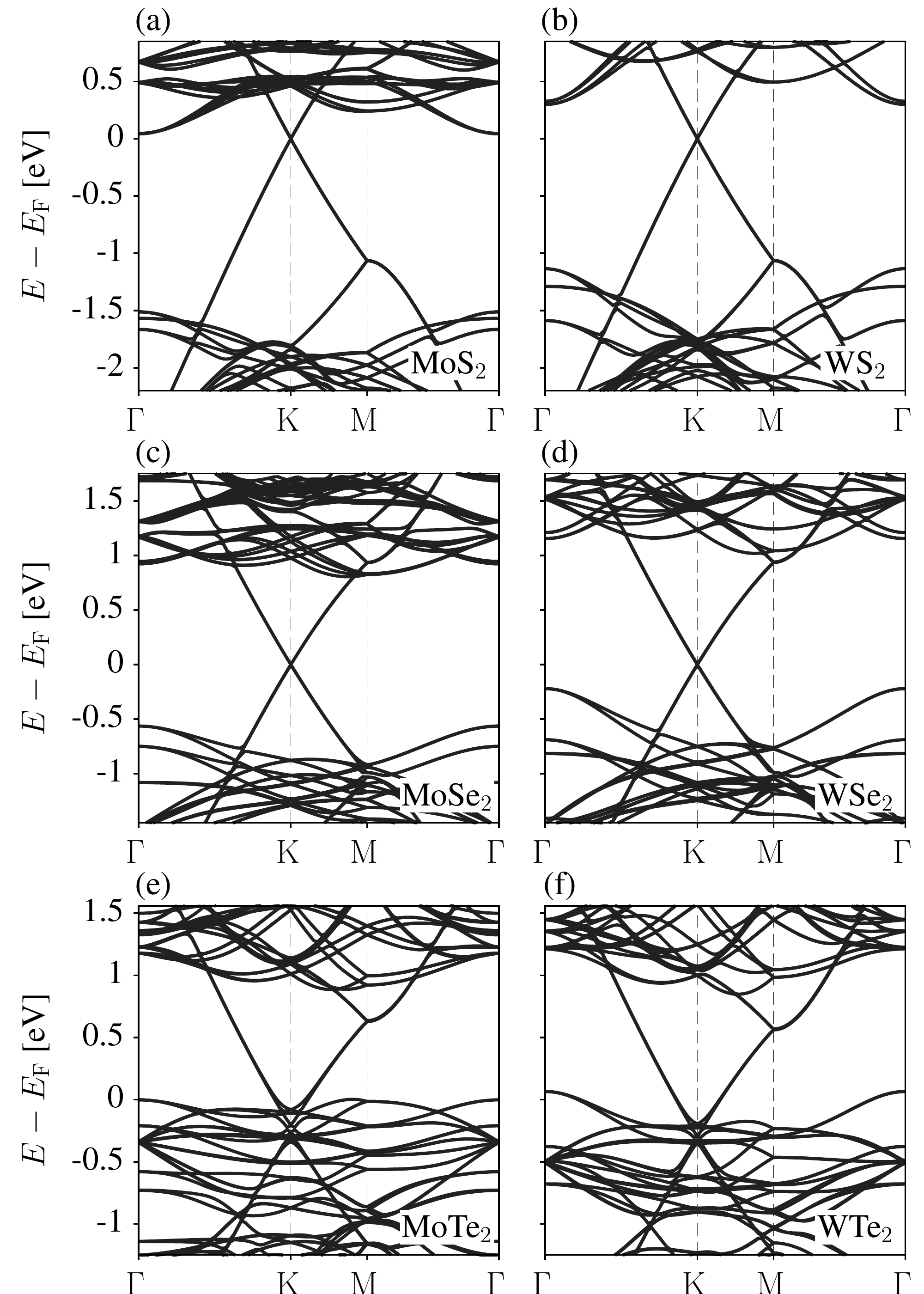}
 \caption{Calculated electronic band structures along high symmetry lines 
for graphene/TMDC heterostructures: (a)~MoS$_2$, (b)~WS$_2$, (c)~MoSe$_2$,
(d)~WSe$_2$, (e)~MoTe$_2$, and (f)~WTe$_2$. }\label{Fig:bands}
\end{figure}

\paragraph{Survey of ab initio band structures of graphene on TMDCs.}
To calculate the electronic structure of graphene on TMDCs we applied density functional theory, coded in {Quantum ESPRESSO}~\cite{Giannozzi2009:JPCM}, on a supercell
structural model to reduce strain due to the incommensurate lattice constants of graphene and TMDCs; see Ref.~\cite{SM} for computational details. Such quasicommensurate superstructures of TMDCs have been grown on HOPG~\cite{Ugeda2014:NatMat}. In Fig.~\ref{Fig:bands} we show the calculated band structures of graphene on monolayer MoS$_2$, WS$_2$,
MoSe$_2$, WSe$_2$, MoTe$_2$, and WTe$_2$ along high symmetry
lines. In the case of sulfur and selenium based TMDCs we find linear dispersive states
of graphene with the Dirac cone within the direct gap of TMDCs. As the atomic
number of the chalcogen increases, the Dirac cone shifts down towards the valence
band edge of TMDCs. In tellurides the Dirac point moves below the valence band edge and
the graphene bands there get strongly distorted.
%------------------------------------------------------------------------
%------------------------------------------------------------------------

In the following we study in detail the electronic states of the well-preserved Dirac
band structures of graphene on sulfides and selenides. Essential calculated orbital
electronic properties, such as the valence and conduction band offsets, $\Delta E_{\rm v}$ and $\Delta E_{\rm c}$, the induced dipole moment (which points towards graphene) of the double-layer structures,
and the work functions $W$ of the graphene and TMDC layers (calculated as the
difference between the self-consistent electrical potential just outside of the layer and
the Fermi level of the whole system), are listed in Tab.~\ref{Tab:all-data}.
We also found that the band offsets can be controlled by an applied transverse
electric field~\cite{SM}. For example, we predict the possibility to tune graphene on WSe$_2$ by gates to reach a {\it massless-massive electron-hole} regime~\cite{SM}. 

%
%------------------------------------------------------------------------
\begin{figure}[h!]
 \includegraphics[width=0.99\columnwidth]{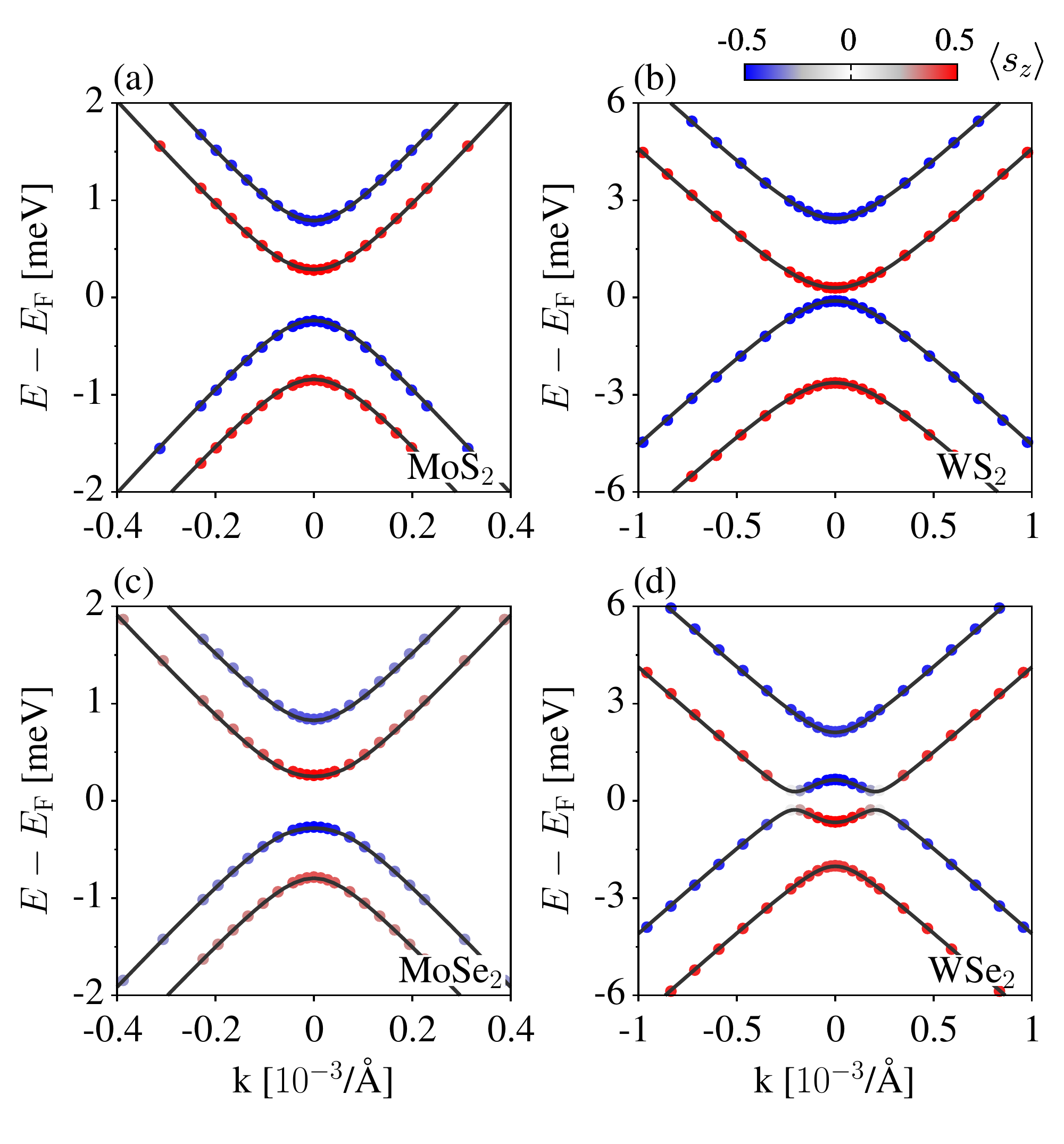}
 \caption{Calculated electronic band structures in the vicinity of the Dirac point for graphene/TMDC heterostructures: (a)~MoS$_2$, (b)~WS$_2$, (c)~MoSe$_2$,
 and (d)~WSe$_2$.
The solid lines are model fits, while the circles are first-principles data. Colors code the
$z$-component of the spin expectation value.
 }\label{Fig:bands_DP}
\end{figure}
%------------------------------------------------------------------------

%------------------------------------------------------------------------
\setlength{\tabcolsep}{6pt} %6pt is default
\begin{table*}

\begin{tabular}{cccccccccccccc}
\hline
 TMDC & $v_{\rm F}/10^5$ & $t$ & $\Delta$ &$\Delta E_{\rm v}$ & $\Delta E_{\rm c}$ & dipole & $W_{\rm grp}$ & $W_{\rm TMDC}$ & $\LIA$ & $\LIB$ & $\LR$ & $\LPIAA$ & $\LPIAB$ \\
      & [m/s] & [eV] & [meV] & [eV] & [eV] & [Debye] & [eV] & [eV] & [meV] & [meV] & [meV] & [meV] & [meV] \\
\hline\hline
% system & vF &   DD   &  DDv &  DDc  & dipole & W_grp & W_TMDC & LIA & LIB & LR & PIA+ & PIA-
 MoS$_2$ & 8.506 & 2.668 & 0.52 & 1.51 & 0.04 & 0.628 & 4.12  & 4.407  & -0.23 & 0.28 & 0.13  & -1.22 & -2.23  \\
MoSe$_2$ & 8.223 & 2.526 & 0.44 & 0.56 & 0.92  & 0.624 & 4.3   & 4.577  & -0.19 & 0.16 & 0.26  & 2.46  & 3.52 \\
  WS$_2$ & 8.463 & 2.657 & 1.31 & 1.13 & 0.30  & 0.675 & 4.12  & 4.432  & -1.02 & 1.21 & 0.36  & -0.98 & -3.81  \\
 WSe$_2$ & 8.156 & 2.507 & 0.54 & 0.22 & 1.15  & 0.641 & 4.3   & 4.587  & -1.22 & 1.16 & 0.56  & -2.69 & -2.54 \\
%MoTe$_2$ & 0   & -108.4& 1258.8& 0.4042 & 0.    & 0.     & 0.     & 0.    & 0.     &  0.    &  0.    \\
% WTe$_2$ & 0   &       &       & 0.     & 0.    & 0.     & 0.     & 0.    & 0.     &  0.    &  0.    \\
\hline
\end{tabular}
\caption{Calculated orbital and spin-orbit parameters, work functions, and dipole moments
for graphene/TMDC heterostructures. Labels: $v_{\rm F}$ is the Fermi velocity of the Dirac states in graphene on TMDCs, $t$ is the hopping energy of graphene's $p_z$ electrons, $\Delta$ is the induced orbital gap of graphene,
 $\Delta E_{\rm v}$ and $\Delta E_{\rm c}$ are the TMDCs
valence and conduction band offsets with respect to graphene's Dirac point, ``dipole'' is the
dipole moment, $W_{\rm grp}$ and $W_{\rm TMDC}$ are the work functions of graphene
and TMDCs, $\LIA$ and $\LIB$ are the intrinsic spin-orbit couplings
for $\mathrm{A}$ and $\mathrm{B}$ graphene sublattices, $\LR$ is the Rashba spin-orbit coupling, and
$\LPIAA$ and $\LPIAB$ are the pseudospin-inversion-asymmetry (PIA)
spin-orbit terms for the two sublattices.
}\label{Tab:all-data}
\end{table*}
%------------------------------------------------------------------------

\paragraph{Dirac band structure topologies.}
We now look at the band structure topologies of the
Dirac cones modified by the proximity effects. Electronic transport in those heterostructures will be graphene-like,
with the proximity-induced fine topological features which depend on the TMDC material.
A zoom into the Dirac cone for the four selected heterostructures is shown in Fig.~\ref{Fig:bands_DP}.
Three materials, graphene on MoS$_2$, WS$_2$, and MoSe$_2$, share the same topology, studied
already in the MoS$_2$ case in Ref.~[\onlinecite{Gmitra:arXiv}]. The essential features are (a) opening
of an orbital gap due to the effective staggered potential (on average, atoms A and B in the
graphene supercell see a different environment coming from the TMDC layer), (b) anticrossing of the
bands due to the intrinsic spin-orbit coupling, and (c) spin splittings of the bands due to spin-orbit coupling
and breaking of the space inversion symmetry. Both the orbital gap and spin-orbit couplings are on the
meV scales, which are giant compared to the 10~$\mu$eV spin-orbit splitting in pristine
graphene~\cite{Gmitra2009:PRB}. In Fig.~\ref{Fig:bands_DP} we also show the spin 
character
of the bands at $\mathrm{K}$. We find that the valence states are formed at the
B sublattice while the conduction states live on A. The same orbital ordering is at $\mathrm{K}'$. The spin alternates as we go through the bands. At $\mathrm{K}'$ the spin orientation is opposite.

The case of graphene on WSe$_2$ stands out. Figure~\ref{Fig:bands_DP} shows an {\it inverted band structure},
which is the main focus of our paper, as it is an indication for a nontrivial topological ordering. While far from
$\mathrm{K}$ the band ordering in the Dirac band structure of WSe$_2$ looks the same as in the other three cases, close
to $\mathrm{K}$ the two lowest energy bands anticross. The top of the valence band and the bottom of the conduction band
have opposite spins to the rest of the states of the same bands.

\paragraph{Effective Hamiltonian.}
Both the trivial and nontrivial topologies observed in Fig.~\ref{Fig:bands_DP} can be modeled with the
same effective Hamiltonian acting on graphene $p_z$ orbitals,
introduced in Ref.~[\onlinecite{Gmitra:arXiv}] for graphene on MoS$_2$.
The Hamiltonian, $H = H_{\rm orb} + H_{\rm so}$ has orbital and spin-orbital parts.
The orbital part, describing gapped Dirac states, is
\begin{equation}
 H_{\rm orb}=\hbar v_{\rm F}(\kappa\sigma_x k_x+\sigma_y k_y) + \Delta\,\sigma_z,
\end{equation}
where $v_{\rm F}$ is the Fermi velocity of Dirac electrons, $\Delta$ is the staggered potential (gap),
$\sigma$ are the pseudospin Pauli
matrices operating on the sublattice $\mathrm{A}$ and $\mathrm{B}$ space, and
$k_x$ and $k_y$ are the Cartesian components of the electron wave vector measured from $\mathrm{K}$\,($\mathrm{K}'$);
parameter $\kappa = 1\,(-1)$ for $\mathrm{K}$\,($\mathrm{K}'$).

The spin-orbit Hamiltonian $H_{\rm so}= H_{\rm I} + H_{\rm R} + H_{\rm PIA}$
has three components: intrinsic, Rashba, and PIA (short for
pseudospin inversion asymmetry~\cite{Gmitra2013:PRL}). Since both intrinsic and PIA are second-nearest
neighbor hoppings~\cite{Konschuh2010:PRB}, they can be different for the two sublattices.
We have,
 \begin{eqnarray}
 H_{\rm I} & = &\frac{1}{2} \left [\LIA  (\sigma_z+\sigma_0) +
\LIB(\sigma_z-\sigma_0)\right ] \kappa s_z, \\
H_{\rm R} & = & \LR(\kappa\sigma_x s_y-\sigma_y s_x), \\
H_{\rm PIA} & = & \frac{a}{2}\left [\LPIAA  (\sigma_z+\sigma_0) +
\LPIAB(\sigma_z-\sigma_0)\right ]\times  \\
& &(k_x s_y - k_y s_x) \nonumber.
 \end{eqnarray}
Here $\LIA$ and $\LIB$ are the intrinsic spin-orbit parameters for sublattice
$\mathrm{A}$ and $\mathrm{B}$, $\LR$ is the strength of the Rashba coupling, and $\LPIAA$ and
$\LPIAB$ are the PIA spin-orbit pa\-ra\-meters; $s$ denotes the spin Pauli matrices,
and $a=2.46$~{\AA} is the pristine graphene lattice constant.

By solving the spectrum of $H$ around $\mathrm{K}$ and comparing with the {\it ab initio} results,
considering the sublattice character of the states as well as their spin projections,
we can uniquely determine the orbital and spin-orbital parameters. They are listed 
in Tab.~\ref{Tab:all-data}. The perfect agreement between the effective model and the
{\it ab initio } calculations, for all four materials, is evident from Fig.~\ref{Fig:bands_DP}.
Both the orbital and spin-orbital parameters can be tuned by a transverse electric field
and vertical strain~\cite{SM}. Only in the case of graphene on WSe$_2$ the orbital gap $\Delta$ is smaller than the magnitudes
of the intrinsic spin-orbit coupling parameters $\lambda_{\rm I}$. This is a signature of the inverted
band structure seen in Fig.~\ref{Fig:bands_DP}.

\paragraph{Quantum spin Hall effect in graphene on WSe$_2$.}
The inverted band structure is a precursor of the quantum spin Hall effect. Although zigzag graphene nanoribbons were predicted to host helical edge states~\cite{Kane2005:PRL}, {\it intrinsic} spin-orbit coupling in graphene is too weak~\cite{Gmitra2009:PRB} for such states to be experimentally realized. Instead, 2d (Hg,Cd)Te quantum wells have emerged as a prototypical quantum spin Hall system~\cite{bernevig_quantum_2006, konig_quantum_2007, zhang_topological_2009}.

%------------------------------------------------------------------------
\begin{figure}[h!]
 \includegraphics[width=0.99\columnwidth]{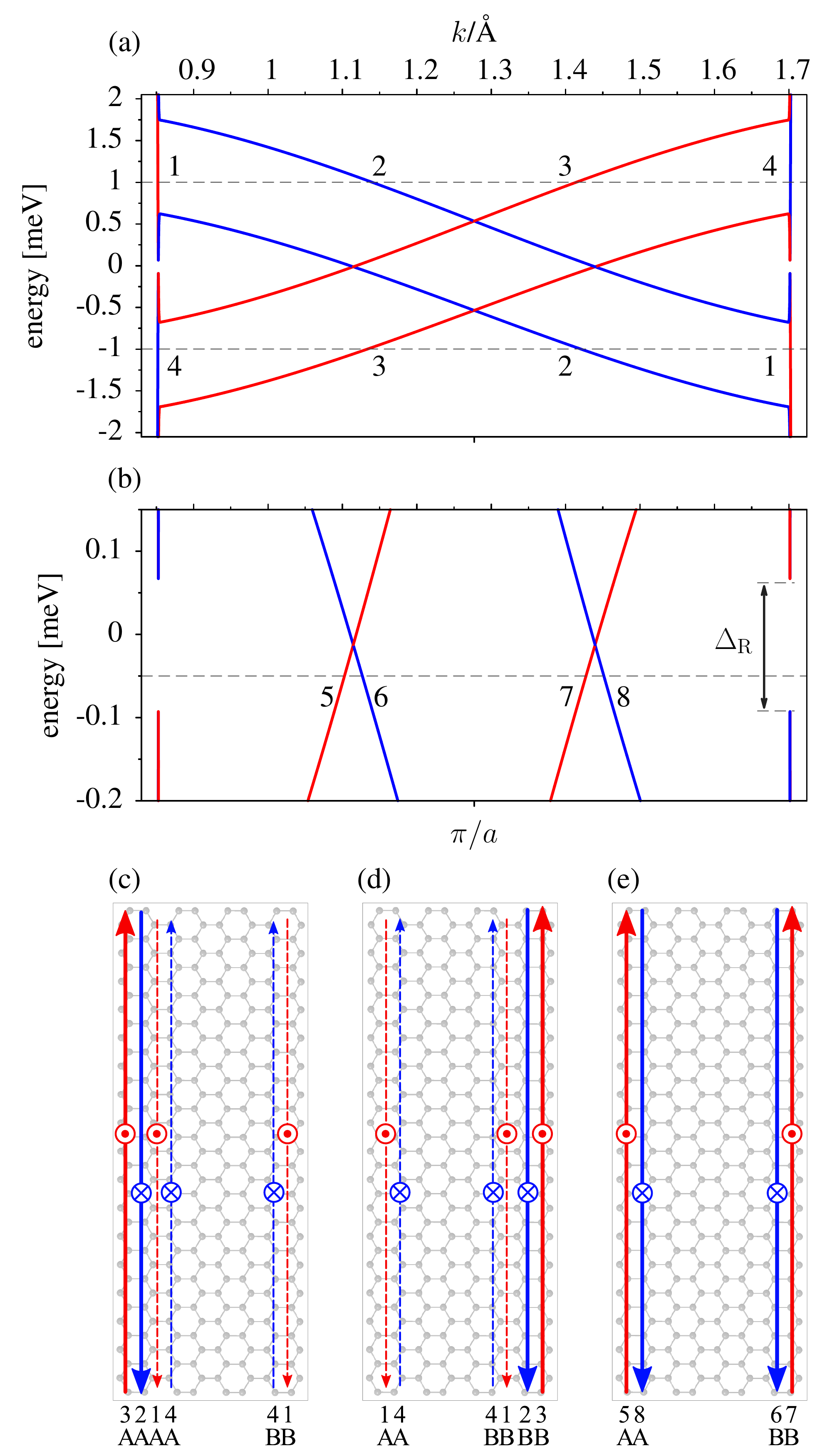}
 \caption{
%Calculated electronic structure of a zigzag graphene nanoribbon on WSe$_2$, with the
%width of 200 nm.
%(a)~Electronic states for spin up (red) and spin down (blue). Labels denote states
%whose localization character is described below in (c)-(e).
%(b)~Zoom to the states within the Rashba anticrossing
% gap $\Delta_{\rm R}$;
%(c)~Sketch of the helical states for energy 1~meV with the labels for
%spatial and pseudospin localization;
%(d)~Sketch of the helical states for energy -1~meV;
%(e)~Sketch of the helical states for energy -0.016~meV;
%Blue (red) arrow corresponds to spin down (up), with respect to the graphene sheet.
%Dashed lines indicate states localized on both edges.
Calculated electronic structure of a zigzag graphene nanoribbon on WSe$_2$, with the
width of 200 nm.
(a)~Electronic states with spin up (red) and spin down (blue). Labels 1-4 denote states
whose localization and sublattice characters are described below in (c)-(e).
(b)~Zoom to the spin polarized states 5-8 within the Rashba anticrossing gap $\Delta_{\rm R}$;
(c)~Sketch of the helical states for energy 1~meV with the labels for
spatial and sublattice localization, as well spin up $\odot$ and spin down $\otimes$ character. 
In-plane vertical up (down) arrows indicate 
positive (negative) group velocities. The dashed lines stand for states localized on both edges;
(d)~Sketch of the helical states for energy -1~meV;
(e)~Sketch of the helical states for energy -0.05~meV inside the Rashba anticrossing gap.}\label{Fig:ribbon}
\end{figure}
%------------------------------------------------------------------------

Our first-principles results strongly suggest that graphene on WSe$_2$, with the inverted Dirac bands due to the strong {\it proximity} (100 times stronger than in pristine graphene) spin-orbit coupling, acts as a quantum spin Hall insulator. In bulk, graphene on monolayer WSe$_2$ experiences a gap, making it an insulator,
see Fig.~\ref{Fig:bands_DP}. This behavior is robust against an applied transverse electric field and vertical
strain~\cite{SM}. To demonstrate the pre\-sence of helical edge states we have converted
our effective Hamiltonian $H$ into a tight-binding model~\cite{SM}, following an earlier work on hydrogenated graphene~\cite{Gmitra2013:PRL}, and analyzed the energy spectra
and states of zigzag nanoribbons of graphene on TMDCs. The results for graphene on WSe$_2$
are shown in Fig.~\ref{Fig:ribbon}, for a nanoribbon of size 200 nm. The band structure
features spin-split bands due to spin-orbit coupling, with four bands crossing the Fermi level. The bulk
gap is transformed to what we term the Rashba anticrossing gap $\Delta_{\rm R}$. This gap increases with the nanoribbon
width as well as with the Rashba coupling, saturating at the bulk level. More details on the Rashba
anticrossing, including perturbative analytical estimates, are presented in Ref.~\cite{SM}, where
we also discuss the offset of the edge-state energies from the bulk states of the nanoribbon.

Already the states above the Rashba
anticrossing gap are peculiar. In Fig.~\ref{Fig:ribbon} we indicate
the states 1-4 at a positive energy of 1~meV. States 1 and 4 are spin-polarized, but localized on both edges.
However, states 2 and 3 are helical, but localized on one edge only! These edge states have a fixed pseudospin character, as shown in the figure. The asymmetry in the edge localization
makes the states 1 and 4 topologically protected against scattering by time reversal impurities at one edge only. At the
other edge backscattering is possible due to the presence of another pair of helical states.
We call such states
half-topological. With increasing width of the
ribbons, states 1 and 4 become more delocalized, eventually becoming bulk states; helical states 2 and 3 stay localized at one edge. At negative energies the asymmetry of the edge states gets reversed, see Fig.~\ref{Fig:ribbon}.

The helical states defining the quantum spin Hall effect live within the Rashba anticrossing
gap $\Delta_{\rm R}$. For example, states 5-8 in Fig.~\ref{Fig:ribbon}(e) are spin-polarized edge states
localized on a specific sublattice as indicated. These states are topologically protected against backscattering by time-reversal impurities. If the model parameters are used for graphene
on MoS$_2$, MoSe$_2$, and WS$_2$, which have trivial Dirac bulk bands, see Fig.~\ref{Fig:bands_DP}, zigzag nanoribbons remain insulating, featuring no helical edge states.

In summary, we have made a detailed study of the electronic states and the proximity spin-orbit
coupling in graphene on monolayer transition-metal dichalcogenides. We have found that graphene
on WSe$_2$ exhibits a band inversion due to spin-orbit coupling. A tight-binding analysis revealed
the presence of half-topological states, protected against backscattering at one edge only, but also
helical edge states, predicting that graphene on WSe$_2$ exhibits the quantum spin Hall
effect.

\begin{acknowledgments}
This work was supported by DFG SFB~689, GRK~1570, International Doctorate Program Topological Insulators of the Elite Network of Bavaria, and by the EU~Seventh~Framework~Programme under Grant~Agreement~No.~604391~Graphene~Flagship.
\end{acknowledgments}

\noindent {\it Note added.} Upon completion of this paper we learned of
a related work by Wang~\emph{et al.}~\cite{morpurgo2015:NComm}.

\bibliography{paper}

\newpage

\begin{widetext}
%------------------------------------------------------------------------
\section*{SUPPLEMENTAL MATERIAL}
%------------------------------------------------------------------------
\end{widetext}

%------------------------------------
\subsection{Computational methods}
\begin{figure}[h!]
  \includegraphics[width=0.8\columnwidth]{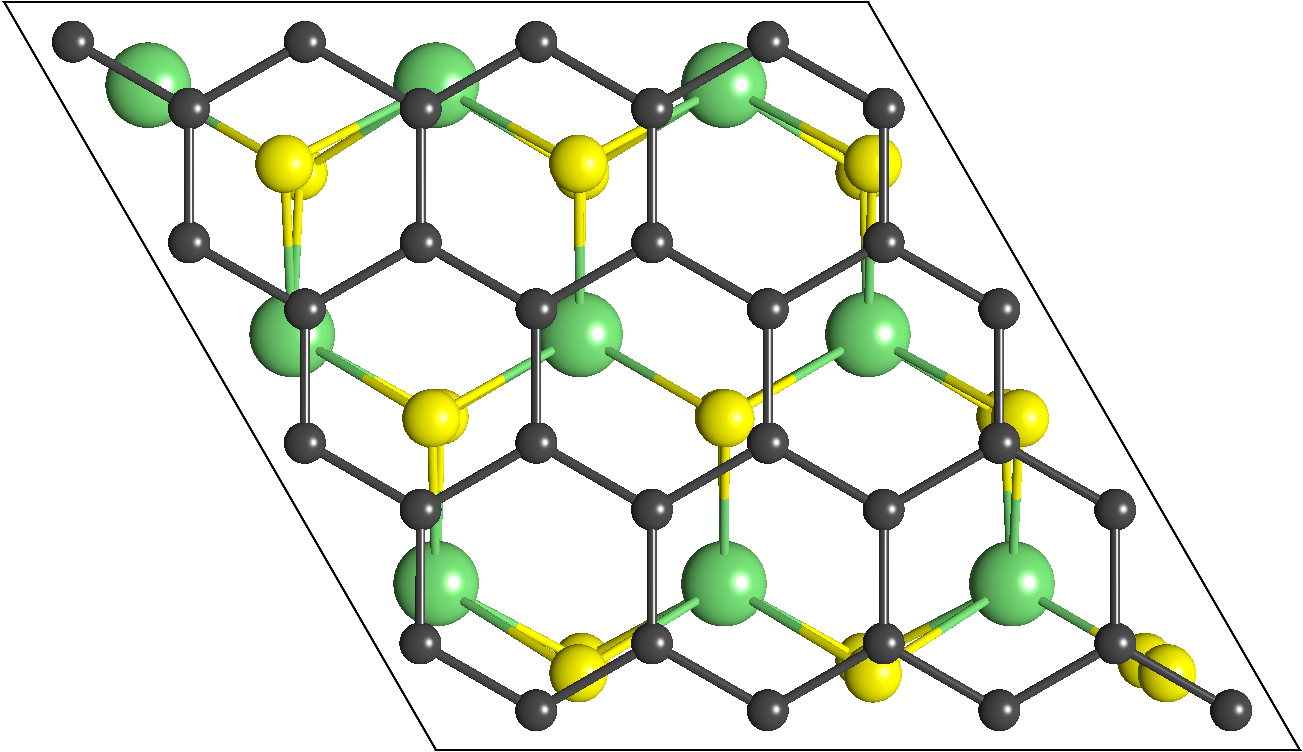}
  \caption{Top view of the structural model of graphene on TMDC used in the DFT calculations.
 The model contains a $4\times 4$ supercell of graphene and a $3\times 3$ supercell of TMDC.
 Carbon atoms are dark gray, metal atoms are green and chalcogen atoms are yellow.
}\label{Fig:struct_model}
\end{figure}

%------------------------------------
%
Structural relaxation and electronic structure
calculations were performed with {Quantum ESPRESSO}~\cite{Giannozzi2009:JPCM},
using norm conserving pseudopotentials with kinetic energy cutoff of 60~Ry for
wavefunctions. For the exchange-correlation potential we used the
generalized gradient approximation~\cite{Perdew1996:PRL}.
To model graphene on TMDC we consider a structural model containing a
$4\times 4$ supercell of graphene and a $3\times 3$ supercell of TMDC,
see Fig.~{\ref{Fig:struct_model}}. The residual lattice mismatch is
split equally between graphene and TMDC.
In Table~\ref{Tab:struct} we give the lattice constants for TMDC
and the residual lateral strain for graphene.
The supercell has 59 atoms.
The reduced Brillouin zone was sampled with $12\times 12$
k points. The atomic positions were relaxed using the quasi-newton algorithm
based on the trust radius procedure including the van der Waals interaction which was treated
within a semiempirical approach~\cite{Grimme2006:JCC,Barone2009:JCC}.
The average graphene surface corrugation calculated from the standard deviation
is listed in Table~\ref{Tab:struct}.
%------------------------------------------------------------------------
%------------------------------------------------------------------------
%------------------------------------------------------------------------
\setlength{\tabcolsep}{7pt} %6pt is default
\begin{table}
\caption{Structural properties of graphene/TMDC hete\-rostructures used in the
DFT calculations, cf.~Fig.~\ref{Fig:struct_model}. We give the lattice constant $a$
of TMDC, the lateral strain of graphene with respect to the unstrained value
for the lattice constant of 2.46~\AA, and
the average surface corrugation of graphene calculated from the standard deviation.
}\label{Tab:struct}
\begin{tabular}{cccc}
\hline
 TMDC & $a$ & strain & corrugation \\
      & [\AA] & [\%] & [pm] \\
\hline\hline
 MoS$_2$ & 3.231  & $-1.5$  &  3.1 \\
MoSe$_2$ & 3.299  & $+0.6$  &  2.2 \\
  WS$_2$ & 3.228  & $-1.6$  &  4.5 \\
 WSe$_2$ & 3.297  & $+0.5$  &  1.8 \\
MoTe$_2$ & 3.407  & $+3.9$  &  1.1 \\
 WTe$_2$ & 3.405  & $+3.8$  &  1.2 \\
\hline
\end{tabular}
\end{table}
%------------------------------------------------------------------------

The supercell was embedded in
a slab geometry with vacuum of about 13~\AA. We applied the dipole correction~\cite{Bengtsson1999:PRB}, which turned out to be crucial to get the numerically accurate Dirac point offsets within TMDC band gap, see Tab.~I in the paper.

%------------------------------------
\subsection{Spin splitting away from $\mathrm{K}$ for graphene on monolayer WSe$_2$}
%------------------------------------

\begin{figure}[h!]
  \includegraphics[width=0.99\columnwidth]{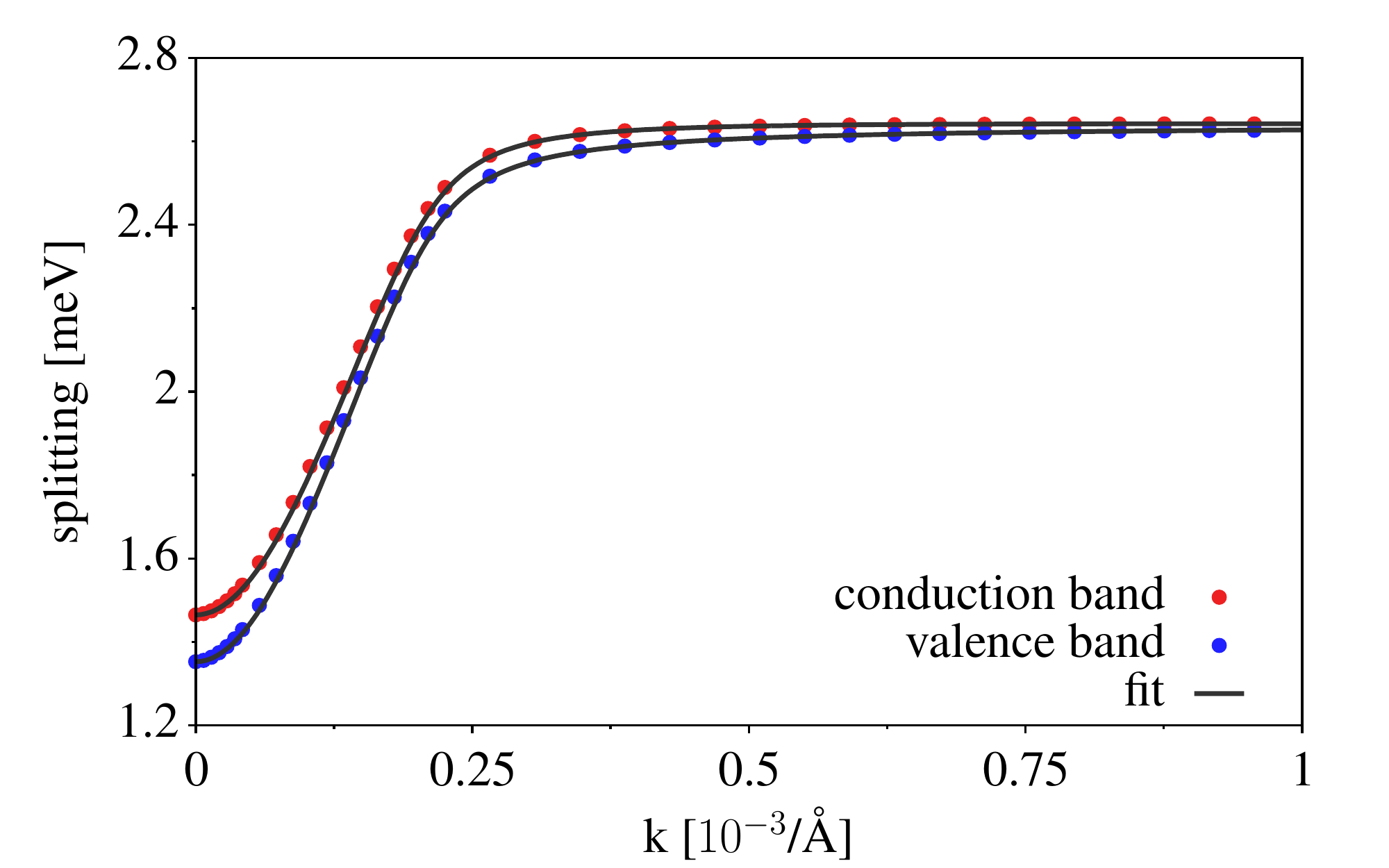}
  \caption{Spin splitting away from $\mathrm{K}$ ($k=0$), in the direction of $\Gamma$, modulated
by PIA spin-orbit coupling in graphene on WSe$_2$.
Solid lines are model fits, symbols are first-principles results.
 }\label{Fig:PIA}
\end{figure}

The pseudospin inversion asymmetry spin-orbit coupling (PIA) is not present directly at $\mathrm{K}$. Away from
$\mathrm{K}$,  PIA introduces a momentum modulation of the spin splitting. In Fig.~\ref{Fig:PIA} we plot the
calculated spin splittings of the valence and conduction bands
of graphene on WSe$_2$. The full effective model Hamiltonian $H$, with PIA, fits the
first-principles data perfectly. The fits give
$\LPIAA=-2.69$~$\mathrm{meV}$ and $\LPIAB=-2.54$~$\mathrm{meV}$. The fitting values,
also for other TMDCs, are presented in Tab.~I in the paper.
%------------------------------------------------------------------------
%------------------------------------------------------------------------

%------------------------------------
\subsection{Effects of transverse electric field and vertical strain for graphene on monolayer WSe$_2$:
robustness of the band inversion}
%------------------------------------

Here we investigate the influence of an applied transverse electric field and vertical strain on the
orbital and spin-orbit parameters for graphene on monolayer WSe$_2$, entering our model Hamiltonian $H$.

\begin{figure}[h!]
  \includegraphics[width=0.99\columnwidth]{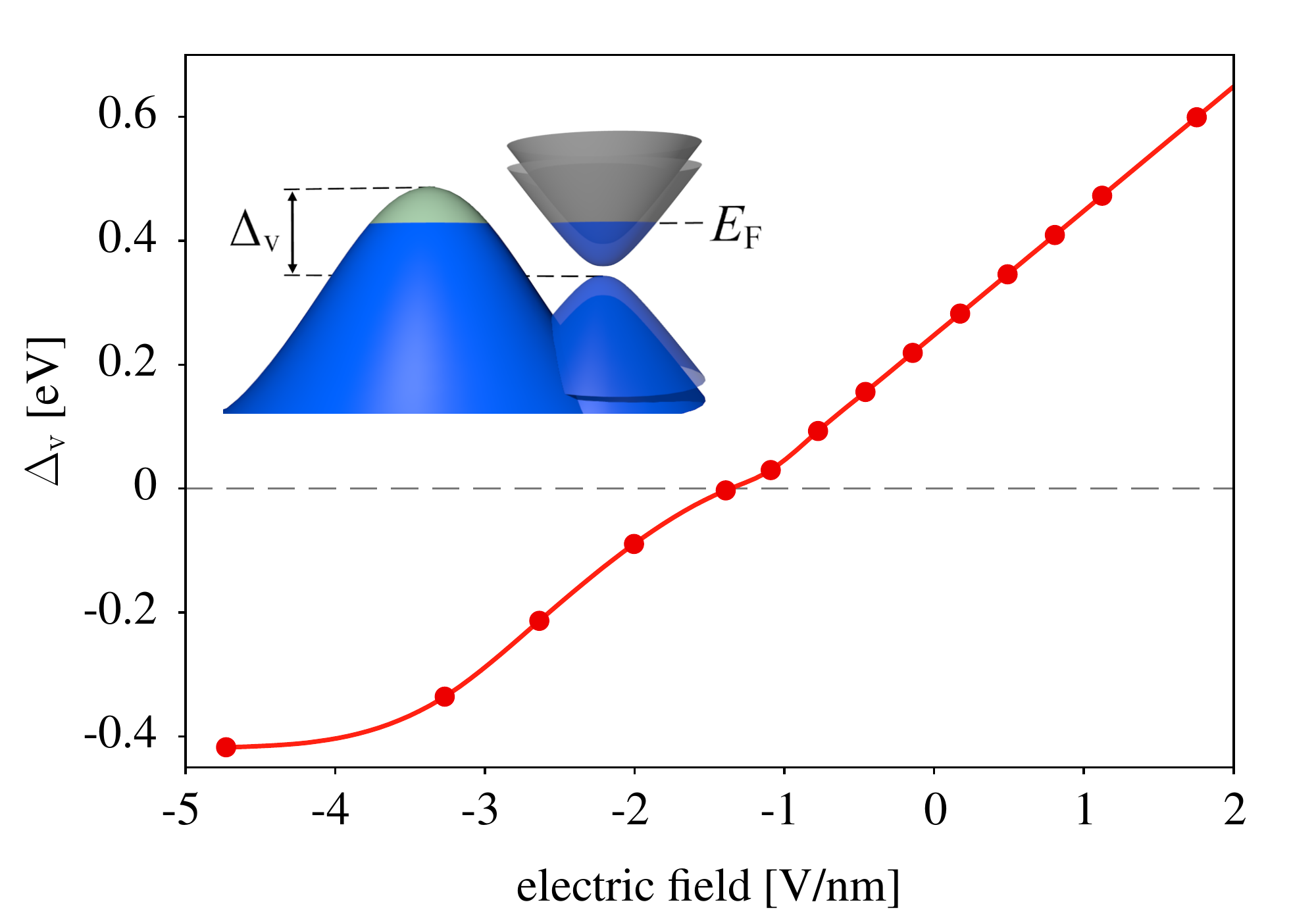}
  \caption{Calculated offset $\Delta_{\rm v}$ from the valence band maximum
of graphene to WSe$_2$, as a function of an applied transverse electric field.
At negative fields electrons are transferred from WSe$_2$ to graphene,
establishing a massless-massive electron-hole bilayer.
}\label{Fig:offset}
\end{figure}

The electric field is included self-consistently on the DFT level. We denote as positive electric fields those pointing
from WSe$_2$ to graphene. Negative fields move the Dirac cone
towards the valence band edge of WSe$_2$.
In Fig.~\ref{Fig:offset} we plot the band offset $\Delta_{\rm v}$, 
which is the
difference between the valence band maximum of graphene and WSe$_2$, as a function of the electric field.
For the fields below -1.4~$\mathrm{V/nm}$ graphene
gets $n$-doped while WSe$_2$ gets $p$-doped. This creates a mixed 
{\it massless-massive electron-hole} system similar to graphene
on MoS$_2$ observed for positive fields~\cite{Gmitra:arXiv}.
%------------------------------------------------------------------------
%------------------------------------------------------------------------

\begin{figure}[h!]
  \includegraphics[width=0.99\columnwidth]{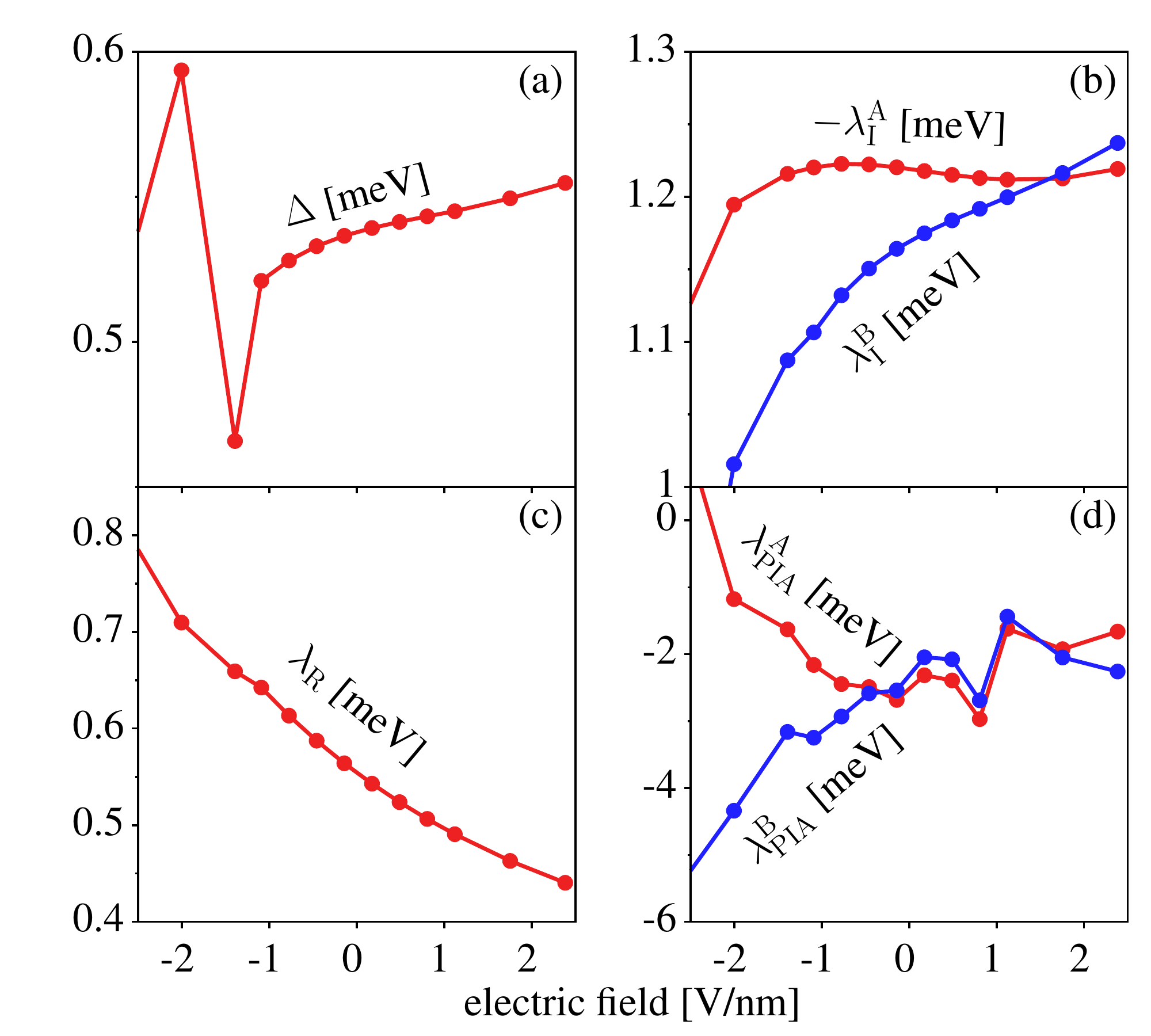}
  \caption{Calculated effective Hamiltonian parameters as a function of transverse electric field
for graphene on WSe$_2$.
(a)~Hybridization gap $\Delta$;
(b)~Sublattice resolved intrinsic spin-orbit coupling $\LIA$ and $\LIB$;
(c)~Rashba parameter $\LR$;
(d)~Pseudospin inversion asymmetry parameters $\LPIAA$ and $\LPIAB$.
 }\label{Fig:efield}
\end{figure}

The effects of the electric field on the Hamiltonian parameters are shown in Fig.~\ref{Fig:efield}.
The orbital gap $\Delta$ does not appreciably change with the field, and similarly the intrinsic spin-orbit couplings
$\lambda_{\rm I}$, which change at most by 20\% in the investigated range of the fields. The Rashba coupling
exhibits a monotonic decay as the electric field increases, changing from 0.8~meV at -2.5~V/nm to 0.45~meV
at 2.5 V/nm. This decrease is appreciable, demonstrating that the Rashba field can be strongly influenced
by the field. The effects on PIA are significant at negative electric fields only. The origin of the observed
dependencies is not obvious. We present them here to show the tunability
of the spin-orbit properties. However, at all the investigated field strengths, graphene on monolayer
WSe$_2$ exhibits the band inversion (this we checked explicitly, but one can also see this
by observing that $\Delta$ is less than the magnitudes of $\lambda_{\rm I}$),
demonstrating its robustness against electric fields, but also the
absence of a possible tunability of the quantum spin Hall effect.
%------------------------------------------------------------------------
%------------------------------------------------------------------------

\begin{figure}[h!]
  \includegraphics[width=0.99\columnwidth]{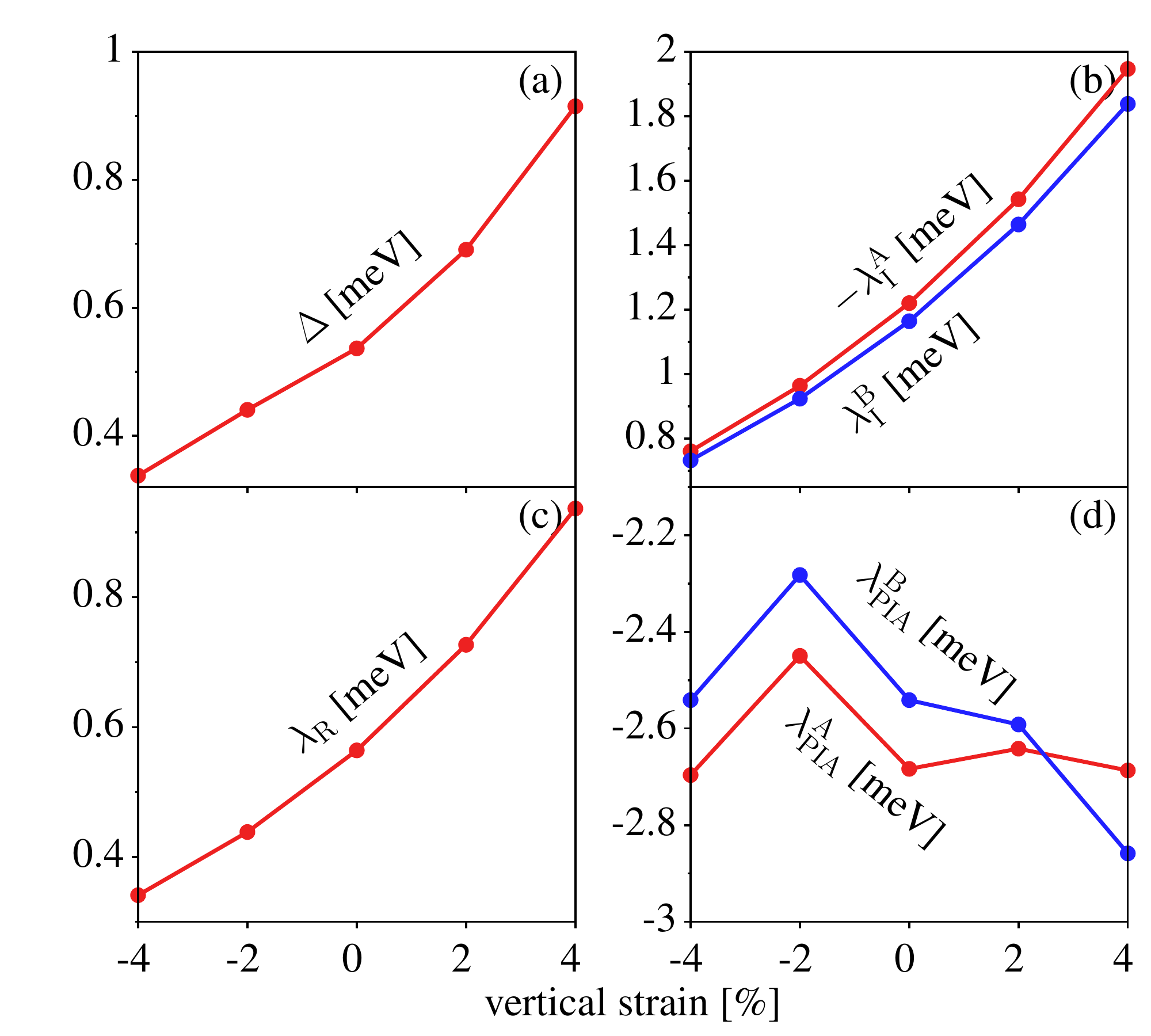}
  \caption{Calculated effective Hamiltonian parameters as in Fig.~\ref{Fig:efield} but
as a function of vertical strain.
}\label{Fig:strain}
\end{figure}

Vertical strain is introduced by changing the interlayer distance between
graphene and WSe$_2$, with respect to the relaxed structure, which is the zero reference strain.
Positive (negative) values of strain correspond to decreased (increased) interlayer distance.
We observe that as the distance between the two layers decreases (strain increases from negative
to positive) the effective model parameters at the K point monotonically increase, with the
exception of the PIA, see Fig.~\ref{Fig:strain}. The increase of the parameters comes
from the increased proximity effects. It is not clear why PIA parameters do not change much
in the investigated regime of strain. But the message, again, is that the band inversion
 is present for all values of the investigated strain, making it robust.

%------------------------------------------------------------------------
%------------------------------------------------------------------------

We conclude that neither an applied transverse electric field, nor a vertical strain
changing the distance of the layers, affect the band inversion predicted for graphene
on monolayer WSe$_2$.

%------------------------------------
\subsection{Effective tight-binding Hamiltonian for graphene on monolayer TMDCs}
%------------------------------------

In the paper we find that the first-principles Dirac band structure of graphene on TMDCs can be
modeled by an effective Hamiltonian $H$ acting on the graphene pseudospin and spin spaces only,
for a given $\mathrm{K}$ ($\mathrm{K}'$). Although the pseudospin symmetry is broken only implicitly, and each
carbon atom in the supercell feels a different local environment, this mapping
of the DFT results on an effective pseudospin-spin Hamiltonian suggests that the effective Hamiltonian
could be also constructed on a tight-binding level.

Indeed, the similarity of $H$ with Hamiltonians with explicit pseudospin symmetry breaking, such as
hydrogenated graphene, allows us to adapt the already derived tight-binding (TB)
Hamiltonian~\cite{Gmitra2013:PRL} to study graphene on monolayer $\mathrm{WSe}_2$. This TB Hamiltonian
extends the graphene Hamiltonian of McClure and Yafet~\cite{j._w._mcclure_theory_1962} and Kane and Mele~\cite{Kane2005:PRL} by adding all symmetry-allowed nearest and next-nearest neighbor terms to
fully maintain the effective sublattice (pseudospin) inversion asymmetry. The Hamiltonian has the form~\cite{Gmitra2013:PRL}:
\begin{eqnarray}\label{Eq:full-TB-inLAOB}
\mathcal{H} &=&
\sum_{\left<i,j\right>} \,t\, c_{i\sigma}^\dagger c^{\phantom\dagger}_{j\sigma}+
\sum_i \,\Delta\, \xi_{c_i}\,c_{i\sigma}^\dagger c^{\phantom\dagger}_{i\sigma} \nonumber \\
&&+\frac{2i}{3}\sum_{\left<i,j\right>}c_{i\sigma}^\dagger c^{\phantom\dagger}_{j\sigma'}\left[\LR\left(\mathbf{\hat{s}}\times \mathbf{d}_{ij}\right)_z\right]_{\sigma\sigma'}\\
&&+\frac{i}{3}\sum_{\left<\left<i,j\right>\right>}c_{i\sigma}^\dagger c^{\phantom\dagger}_{j\sigma'} \left[\frac{\lambda_{\rm I}^{c_i}}{\sqrt{3}}\nu_{ij}\hat{s}_z + 2\lambda_{\mathrm{PIA}}^{c_i} \left(\mathbf{\hat{s}}\times \mathbf{D}_{ij}\right)_z\right]_{\sigma\sigma'}\,.\nonumber
\end{eqnarray}
where $c_{i\sigma}^\dagger=\left(a_{i\sigma}^\dagger,b_{i\sigma}^\dagger\right)$ and $c_{i\sigma}=\left(a_{i\sigma},b_{i\sigma}\right)$ denote the creation and annihilation operators for an electron on a lattice site $i$ that belongs to the sublattice A or B, respectively, and hosts spin $\sigma$. The first two terms in Eq.~(\ref{Eq:full-TB-inLAOB}) govern dynamics on the orbital energy scale; the nearest neighbor hopping (sum over $\left<i,j\right>$) is parameterized by a hybridization $t$, and the staggered on-site potential $\Delta$ accounts for an effective energy difference experienced by atoms in the sublattice A ($\xi_{a_i}=1$) and B ($\xi_{b_i}=-1$), respectively. The three remaining terms in $\mathcal{H}$, Eq.~(\ref{Eq:full-TB-inLAOB}), describe spin-orbit coupling (SOC) via the nearest (sum over $\left<i,j\right>$) and next-nearest (sum over $\left<\left<i,j\right>\right>$) neighbor hoppings. The first of the last three terms is the Rashba SOC parameterized by $\LR$. It arises because the inversion symmetry is broken when graphene is placed on top of $\mathrm{WSe}_2$. The last two next-nearest neighbor terms in Eq.~(\ref{Eq:full-TB-inLAOB}) are the sublattice resolved
intrinsic, $\lambda_{\rm I}^{c_i}=\LIA(\LIB)$ for $c_i$ on sublattice A(B), and the pseudospin inversion asymmetry (PIA) induced term parameterized by $\lambda_{\mathrm{PIA}}^{c_i}=\LPIAA(\LPIAB)$ for $c_i$ on sublattice A(B), respectively.
Both terms appear since the sublattice (pseudospin) symmetry is broken on average. Here, $\mathbf{\hat{s}}$ is a vector of Pauli matrices acting on the spin space and the sign factor $\nu_{ij}=1(-1)$ stands for the clockwise (counterclockwise) hopping path from site $j$ to site $i$. The unit vectors pointing from site $j$ to site $i$ are denoted by $\mathbf{d}_{ij}$ for the nearest neighbors, and by $\mathbf{D}_{ij}$ for the next-nearest neighbors.

%------------------------------------
\subsection{Tight-binding Hamiltonian in the Bloch basis.}
%------------------------------------

To calculate the energy spectrum 
%of a nanoribbon on $\mathrm{WSe}_2$ we assume periodicity along the direction parallel with the zigzag edges and we perform a partial Fourier transformation of %the local atomic creation and annihilation operators along this direction. 
we rewrite the ori\-ginal tight-binding Hamiltonian $\mathcal{H}$, Eq.~(\ref{Eq:full-TB-inLAOB}), via the associated Bloch state operators $c^\dagger_{\sigma}(\mathbf{q})$ and $c_{\sigma}(\mathbf{q})$ defined as follows:
\begin{align}
c^\dagger_\sigma(\mathbf{q})&=\frac{1}{\sqrt{N}}\sum\limits_{m}e^{i\mathbf{q}\cdot\mathbf{R}_m}\,c^\dagger_{m,\sigma}\,, &\\
c_\sigma(\mathbf{q})&=\frac{1}{\sqrt{N}}\sum\limits_{m}e^{-i\mathbf{q}\cdot\mathbf{R}_m}\,c_{m,\sigma}\,,
\end{align}
where $\mathbf{R}_m$ is the lattice vector of an atomic site $m$, and $m$ runs over all $N$ atomic sites (in the given sublattice) forming the macroscopic system. After the transformation, $\mathcal{H}=\sum_{\mathbf{q}} H(\mathbf{q})$, where the particular Bloch Hamiltonian $H(\mathbf{q})$ as expressed in the ordered
Bloch basis, $\bigl\{a_\uparrow(\mathbf{q}),a_\downarrow(\mathbf{q}),b_\uparrow(\mathbf{q}),b_\downarrow(\mathbf{q})\bigr\}$, reads:
\begin{widetext}
\begin{equation}
H(\mathbf{q})=
\left(
\begin{array}{cccc}
\Delta -\lambda_{\rm I}^{\rm A}\,f_{\rm I}(\mathbf{q})                      &
\lambda_{\rm PIA}^{\rm A}\, f_{\rm PIA}(\mathbf{q})                           &
t\, f_{\rm orb}(\mathbf{q})                                                 &
i\lambda_{\rm R}\,f_{\rm R}(\mathbf{q})                                     \\
%%%
\lambda_{\rm PIA}^{\rm A}\,f_{\rm PIA}^*(\mathbf{q})                 &
\Delta +\lambda_{\rm I}^{\rm A}\,f_{\rm I}(\mathbf{q})                     &
i\lambda_{\rm R}\,f_{\rm R}^*(-\mathbf{q})                         &
t\,f_{\rm orb}(\mathbf{q})                                                  \\
%%%
t\,f_{\rm orb}^*(\mathbf{q})                                       &
-i\lambda_{\rm R}\,f_{\rm R}(-\mathbf{q})                                   &
-\Delta +\lambda_{\rm I}^{\rm B}\,f_{\rm I}(\mathbf{q})                    &
-\lambda_{\rm PIA}^{\rm B}\,f_{\rm PIA}(\mathbf{q})                           \\
%%%
-i\lambda_{\rm R}\,f_{\rm R}^*(\mathbf{q})                         &
t\,f_{\rm orb}^*(\mathbf{q})                                       &
-\lambda_{\rm PIA}^{\rm B}\,f_{\rm PIA}^*(\mathbf{q})                &
-\Delta -\lambda_{\rm I}^{\rm B}\,f_{\rm I}(\mathbf{q})
\end{array}
\right)\,.
\end{equation}
The orbital and spin-orbital structural tight-binding functions $f_{\rm orb}, f_{\rm I}, f_{\rm R}, f_{\rm PIA}$ are defined as follows:
\begin{align}
f_{\rm orb}(\mathbf{q}) &= 1 + 2 e^{i\frac{\sqrt{3}}{2} q_y a}\cos{q_x a}, &
f_{\rm R}(\mathbf{q})  &=\frac{2}{3}\Bigl[1 + e^{-i\frac{2\pi}{3}} e^{\frac{i}{2}(q_x + \sqrt{3} q_y)a} +  e^{i\frac{2\pi}{3}} e^{\frac{i}{2}(-q_x + \sqrt{3} q_y)a}\Bigr],\\
f_{\rm I}(\mathbf{q})  &=\frac{4}{3\sqrt{3}} \Big[\cos{\tfrac{\sqrt{3} q_y a}{2}} - \cos{\tfrac{q_x a}{2}} \Bigr]\sin{\tfrac{q_x a}{2}}, &
%f_{\rm I}(\mathbf{q})  &=\frac{2}{3\sqrt{3}} (-\sin{q_x a} + \sin{\tfrac{q_x-\sqrt{3} qy}{2}a} + \sin{\tfrac{q_x+\sqrt{3} qy}{2}a}), &
f_{\rm PIA}(\mathbf{q})  &=\frac{4}{3}i\Bigl[\cos{\tfrac{\sqrt{3} q_y a}{2}} \sin{\tfrac{q_x a}{2}} + \sin{q_x a} -
   i \sqrt{3} \cos{\tfrac{q_x a}{2}}\sin{\tfrac{\sqrt{3} q_y a}{2}}\Bigr],
\end{align}
where $a=2.46$~\AA, and $q_x$ and $q_y$ are the Cartesian components of the $\mathbf{q}$-vector with respect to the center of the Brillouin zone ($\Gamma$). The low-energy physics near the given valley $\kappa\mathrm{K}$ can be effectively described by the Hamiltonian $H(\kappa\mathrm{K}+\mathbf{k})$ expanded in $\mathbf{k}$ to the first order, keeping for each coupling constant only the leading term in the $\mathbf{k}$-expansion. For example at the $\mathrm{K}$ valley we get:
\begin{equation}
H_\mathrm{K}^{\mathrm{eff}}(\mathbf{k})
=\left(
\begin{array}{cccc}
\Delta +\lambda_{\rm I}^{\rm A}         & a \lambda_{\rm PIA}^{\rm A} (-i k_x - k_y) & a\tfrac{\sqrt{3}}{2} t (k_x - i k_y)       & 0 \\
a \lambda_{\rm PIA}^{\rm A}(i k_x - k_y)& \Delta-\lambda_{\rm I}^{\rm A}             & 2 i \lambda_{\rm R}                        & a\tfrac{\sqrt{3}}{2} t (k_x - i k_y) \\
a \tfrac{\sqrt{3}}{2} t (k_x + i k_y)   & -2 i \lambda_{\rm R}                       & -\Delta-\lambda_{\rm I}^{\rm B}            & a\lambda_{\rm PIA}^{\rm B} (i k_x + k_y) \\
 0                                      & a \tfrac{\sqrt{3}}{2} t (k_x + i k_y)      & a \lambda_{\rm PIA}^{\rm B} (-i k_x + k_y) & -\Delta+\lambda_{\rm I}^{\rm B}
\end{array}
\right)=H_{\mathrm{orb}}+H_{\mathrm{I}}+H_{\mathrm{R}}+H_{\mathrm{PIA}}\,,
\end{equation}
where $H_{\mathrm{orb}},H_{\mathrm{I}},H_{\mathrm{R}}$ and $H_{\mathrm{PIA}}$ are explicitly written in the paper, see Eqs.~(1-4), with the velocity 
$v_{\rm F}=\tfrac{\sqrt{3}}{2}a t/\hbar$. This demonstrates the
consistency of the effective Hamiltonian $H$ in the paper and the TB Hamiltonian described here.
\end{widetext}

\begin{figure}[h!]
  \includegraphics[width=0.99\columnwidth]{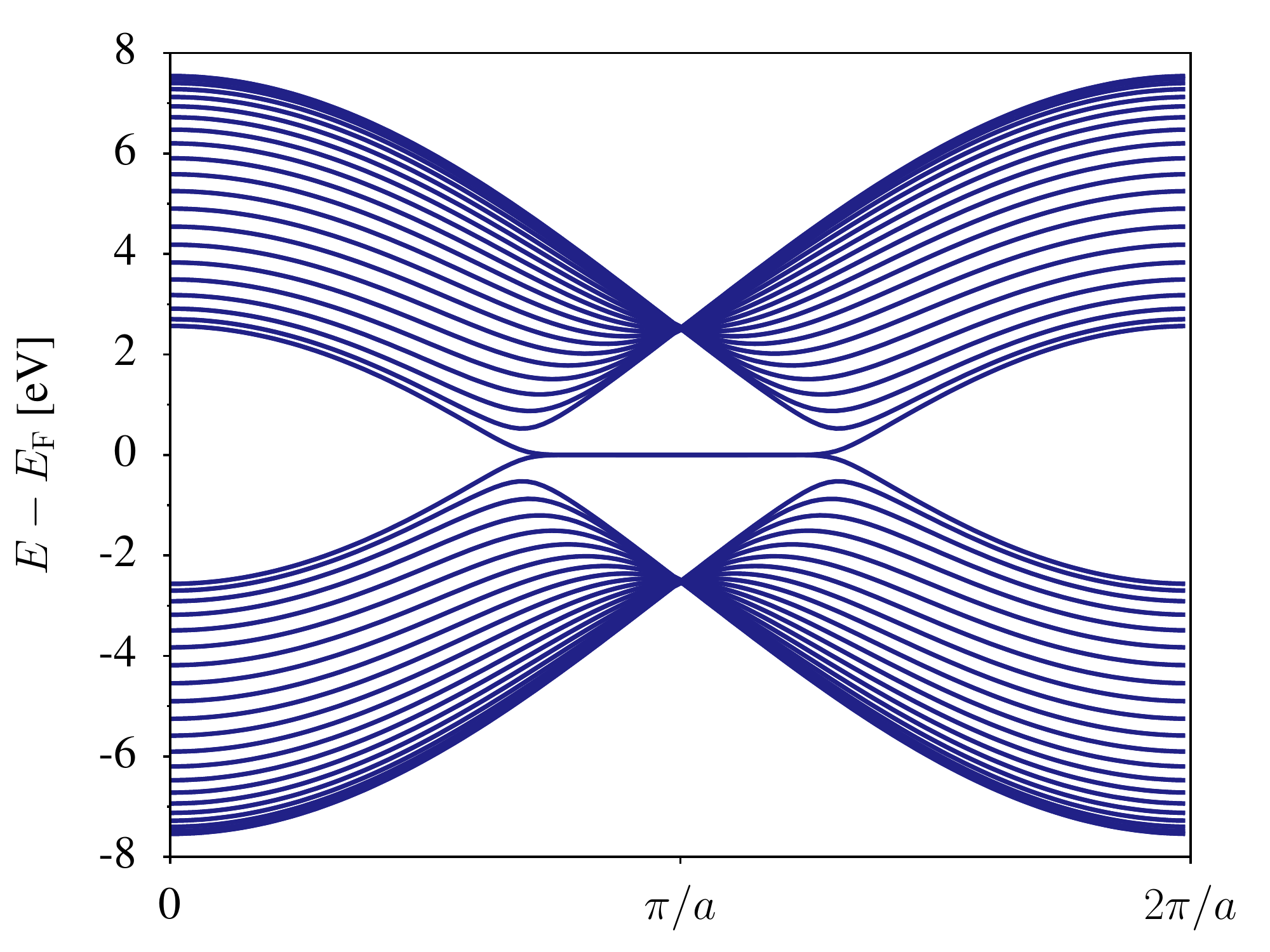}
  \caption{Calculated band structure for a zigzag graphene nanoribbon 
  on WSe$_2$ with the width of 4.3~nm. Relevant tight-binding parameters used
  in the calculation are from Tab.~I in the paper.
 }\label{Fig:tb-bands}
\end{figure}

The parameters for the TB Hamiltonian are included in Tab.~I of the paper.
The calculated electronic band structure for a zigzag nanoribbon of graphene on WSe$_2$, of 4.3 nm width, 
is shown in Fig.~\ref{Fig:tb-bands}. Zooms of such a band structure at the region around the Fermi level 
are in Fig.~4(a) of the paper, for a wider ribbon.
%------------------------------------------------------------------------
%------------------------------------------------------------------------

%------------------------------------
\subsection{Rashba anticrossing gap $\Delta_{\rm R}$ and its wave vector $k_{\rm R}$.}
%------------------------------------

As discussed in the paper, the helical edge states live inside the Rashba anticrossing gap $\Delta_{\rm R}$. Here we
study this gap with our tight-binding model, and provide analytical formulas which demonstrate clearly
its origin.

\begin{figure}[h!]
  \includegraphics[width=0.99\columnwidth]{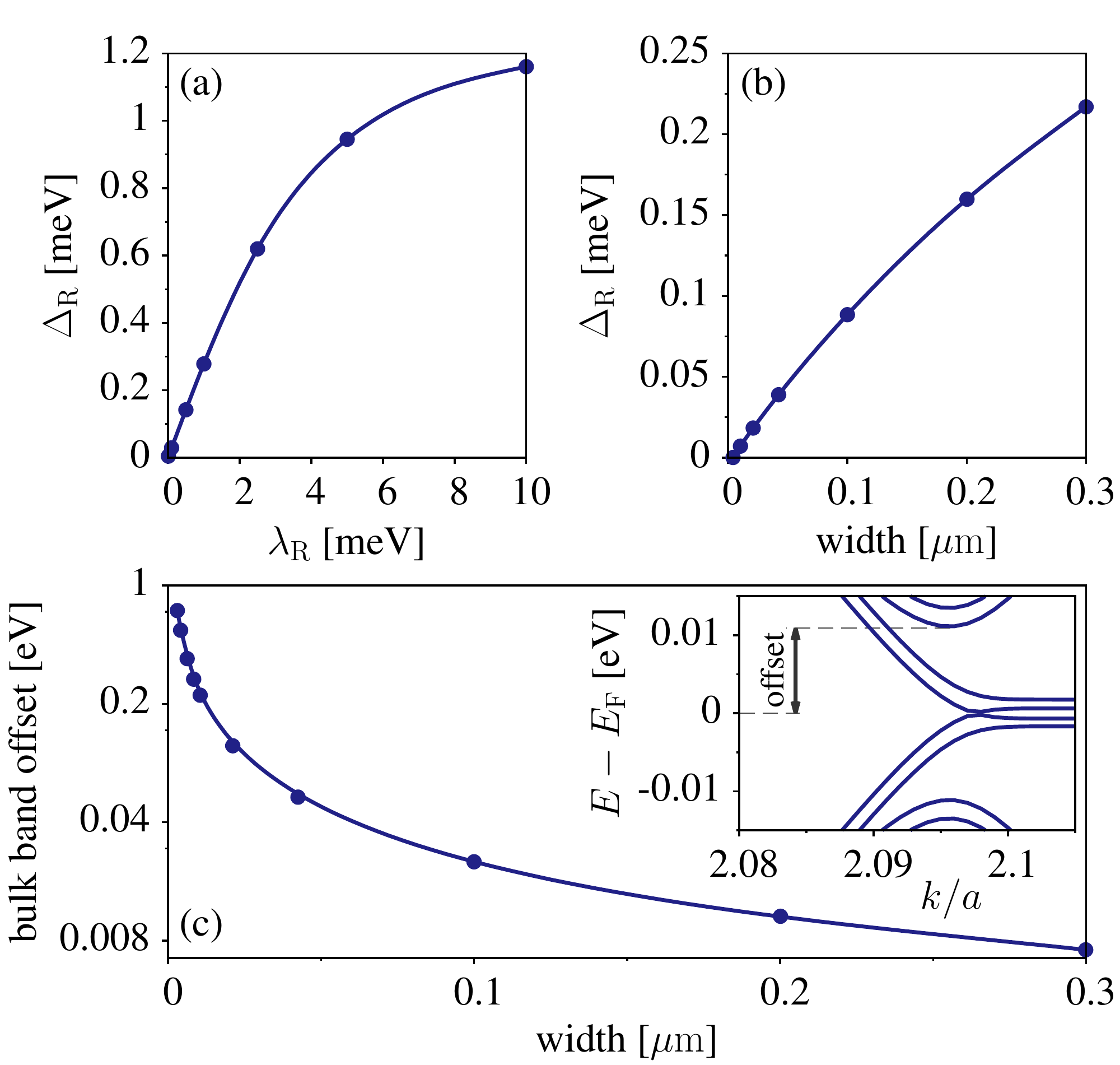}
  \caption{Calculation of the Rashba anticrossing gap $\Delta_{\rm R}$ for a zigzag graphene nanoribbon on WSe$_2$
(a)~as a function of Rashba SOC strength for a fixed narrow nanoribbon of 200~nm width,
(b)~as a function of the nanoribbon width for a fixed $\LR=0.56$~meV.
(c)~Energy offset of the nanoribbon states with a bulk character (see inset) as a function of nanoribbon width. 
Model parameters are provided in Tab.~I of the paper.
 }\label{Fig:tb-gap}
\end{figure}

In Fig.~\ref{Fig:tb-gap}(a) we plot $\Delta_{\rm R}$ as a function of the Rashba SOC parameter
$\LR$ for a narrow nanoribbon of 200~nm.
The gap increases linearly with
$\LR$ for small, but physically relevant $\LR$, then it starts to saturate.
Also, for the $\LR$ parameter of graphene on WSe$_2$, the Rashba anticrossing gap
increases linearly with increasing nanoribbon width, see
Fig.~\ref{Fig:tb-gap}(b) expected to reach the bulk gap of about 0.56~meV at large widths.

We also looked at the offset between the bulk and edge nanoribbon states. The results are shown
in Fig.~\ref{Fig:tb-gap}(c). As the nanoribbon width increases, the bulk states move closer to the
zero energy level. For a relatively wide nanoribbon of 0.3~$\mu$m, the Rashba gap $\Delta_{\rm R}$=0.2~meV and
the bulk band offset is 7~meV.
%------------------------------------------------------------------------
%------------------------------------------------------------------------

Finally, we give analytical estimates of the Rashba anticrossing energy $\Delta_{\rm R}$ and the wave vector
$k_{\rm R}$ at which the anticrossing occurs. These are the main characteristics of the
\emph{inverted band structure}. To this end we analyze the spectrum of $H_\mathrm{K}^{\mathrm{eff}}(\mathbf{k})$. We consider $H_{\mathrm{orb}}+H_{\mathrm{I}}$ as the unperturbed Hamiltonian and treat $H_{\mathrm{R}}$ as a perturbation. We neglect the PIA Hamiltonian $H_{\mathrm{PIA}}$ because of its
$\mathbf{k}$-dependence near the center of the $\kappa\mathrm{K}$ valley; the effects of $\LPIA$
there are much weaker than that of $\LR$.
The eigenspectrum of $H_{\mathrm{orb}}+H_{\mathrm{I}}$ reads:
\begin{equation}
E_{\pm,\pm}(\mathbf{k})=
\pm\tfrac{\lambda_{\rm I}^{\rm A}-\lambda_{\rm I}^{\rm B}}{2}
\pm\sqrt{\tfrac{3t^2 a^2|\mathbf{k}|^2}{4}+\left(\Delta-\tfrac{\lambda_{\rm I}^{\rm A}+\lambda_{\rm I}^{\rm B}}{2}\right)^2}\,.
\end{equation}
Depending on the relative signs and magnitudes of $\Delta$, $\LIA$ and $\LIB$, two bands out of four
$E_{\pm,\pm}(\mathbf{k})$ always cross. The momenta where this crossing happens form, around each valley center,
a circle with radius $k_{\mathrm{R}}$.
In our representative case corresponding to graphene on WSe$_2$ the magnitudes of the relevant parameters are ordered as $0<\Delta<\lambda_{\rm I}^{\rm B}<-\lambda_{\rm I}^{\rm A}$,
and in this configuration the bands $E_{+,+}(|\mathbf{k}|)$ and $E_{-,-}(|\mathbf{k}|)$ cross at
\begin{equation}
k_{\mathrm{R}}=\frac{2}{\sqrt{3} t a}\sqrt{(\Delta-\lambda_{\rm I}^{\rm A})(\lambda_{\rm I}^{\rm B}-\Delta)}\,.
\end{equation}
The perturbation $H_{\mathrm{R}}$ removes the degeneracy along the $k_{\mathrm{R}}$-circle and opens a gap $\Delta_{\rm R}$. Treating $H_{\mathrm{R}}$ within the first order perturbation theory for degenerate spectra we
obtain,
\begin{widetext}
\begin{equation}
\Delta_{\rm R}=E_{-,-}^{\rm pert}(k_{\mathrm{R}})-E_{+,+}^{\rm pert}(k_{\mathrm{R}})=4\lambda_{\rm R}\sqrt{\frac{\lambda_{\rm I}^{\rm B}-\Delta}{\lambda_{\rm I}^{\rm B}-\lambda_{\rm I}^{\rm A}}}
\left[ 1+\frac{(\lambda_{\rm I}^{\rm B}+\Delta)^2}{(\lambda_{\rm I}^{\rm B}-\Delta)(\Delta-\lambda_{\rm I}^{\rm A})} \right]^{-1/2}.
\end{equation}
\end{widetext}
Plugging for the staggered potential $\Delta$ and SOC strengths $\LIA$, $\LIB$ and $\LR$ 
from the Tab.~I of the paper, we get for graphene on WSe$_2$
$k_{\rm R}=0.2\,[10^{-3}/\text{\AA}]$ and $\Delta_{\rm R}=0.6$~meV. These values 
are a very good approximation to the computed DFT-characteristics of 
the inverted band structure as seen at Fig.~3(d) in the paper.

\end{document}